\documentclass[aps,twocolumn,superscriptaddress,showpacs]{revtex4}
\usepackage{graphicx}

\begin{document}
\title{Two-peaked and flat-top perfect bright solitons in epsilon-near-zero nonlinear metamaterials: novel Kerr self-trapping mechanisms}

\author{C. Rizza}
\affiliation{Dipartimento di Ingegneria Elettrica e dell'Informazione, Universit\`{a} dell'Aquila 67100, Monteluco di Roio, Italy}
\author{A. Ciattoni}
\email{alessandro.ciattoni@aquila.infn.it} \affiliation{Consiglio Nazionale delle Ricerche, CNR-SPIN, 67100 L'Aquila, Italy} \affiliation{Dipartimento di
Fisica, Universit\`{a} dell'Aquila 67100, Italy}
\author{E. Palange}
\affiliation{Dipartimento di Ingegneria Elettrica e dell'Informazione, Universit\`{a} dell'Aquila 67100, Monteluco di Roio, Italy}
\date{\today}

\begin{abstract}
We analytically investigate transverse magnetic (TM) spatial bright solitons, as exact solutions of Maxwell's equations, propagating through
nonlinear metamaterials whose linear dielectric permittivity is very close to zero and whose effective nonlinear Kerr parameters can be tailored to
achieve values not available in standard materials. Exploiting the fact that, in the considered medium, linear and nonlinear polarization can be
comparable at feasible and realistic optical intensities, we identify two novel self-trapping mechanisms able to support two-peaked and flat-top
solitons, respectively. Specifically, these two novel mechanisms are based on the occurrence of critical points at which the effective nonlinear
permittivity vanishes, the two mechanisms differing in the way the compensation between linear and nonlinear polarization is achieved through the
non-standard values of the nonlinear parameters.
\end{abstract}

\maketitle

\section{Introduction}

Metamaterials have recently attracted a large attention since they are characterized by unconventional electromagnetic properties and have recently
suggested novel and fascinating applications \cite{Pendry1,Fang,Pendry2,Schurig,Chen}. In the context of nonlinear optics, metamaterials have
suggested two different ways to overcome its fundamental limit consisting in the fact that, due to the weakness of the nonlinear response, nonlinear
behavior is observed at large optical intensities. In the first scheme the material is tailored in such a way that the nonlinear medium is placed at
the region where a large field enhancement occurs \cite{Fischer,Pendry3}, whereas in the second scheme the linear dielectric permittivity is chosen
to be so small to allow the nonlinear polarization to fully rule the electromagnetic behavior (\emph{extreme nonlinear regime}) \cite{Ciatt1}. Such
an extreme nonlinear regime has been recently allowed to predict genuinely novel and intriguing phenomena as, for example, beam transverse power flow
reversing \cite{Ciatt2} where Poynting vector flips its sign along the transverse profile of an optical beam since optical radiation nonlinearly
produces a region (around the beam propagation axis) where the dielectric behaves as a metal. A nonlinear slab characterized by a very small optical
intensity has been shown to support transmissivity directional hysteresis \cite{Ciatt3} and the corresponding underlying electromagnetic multiplicity
has been shown to be produced by the nonlinear field matching at the slab interfaces in the extreme nonlinear regime. In the context of
nondiffracting and self-trapped waves, nonlinear epsilon-near-zero metamaterials have been recently shown to be able to support dark solitons
characterized by a central and extended region where Poynting vector exactly vanishes together with the transverse electric field component
\cite{Rizza1}.

In this paper, we analytically investigate transverse magnetic (TM) spatial bright solitons propagating through epsilon-near-zero nonlinear Kerr
metamaterials (material proposed in Ref.\cite{Ciatt1}) and we predict the existence of two novel perfect soliton \cite{Ciatt4} families whose
Poynting vector profiles are two-peaked and flat-top \cite{Goriii,Maaaaa}, respectively. Such metamaterials host the extreme nonlinear regime where,
due to the small value of the linear dielectric permittivity, linear and nonlinear polarizations can be comparable at feasible optical intensities.
As a consequence, critical electromagnetic points are allowed where the overall effective nonlinear permittivity vanishes, and in this paper we prove
that these points can be profitably exploited to tailor the soliton shape, allowing us to obtain both two-peaked and flat-top solitons. The
occurrence of such points literally provides novel physical self-trapping mechanisms (much more involved than standard compensation between
self-focusing and diffraction) where the longitudinal electric field component (i.e. the component parallel to the soliton propagation direction) is
the basic mean to produce the above mentioned balance between linear and nonlinear polarizations. The full exploitation of such novel self-trapping
mechanisms is possible since the nonlinear parameters of the metamaterial Kerr response (as discussed in Ref.\cite{Ciatt1}) can be tailored to assume
values not available in standard materials. In order to prove the feasibility of the two novel soliton families, we propose a realistic composite
medium characterized by a sufficient number of parameters which can be independently tuned to allow the overall homogenized medium response to meet
the conditions required for observing both two-peaked and flat-top solitons. In addition, as a consequence of the small value of the linear
permittivity, two-peaked and flat-top solitons turn out to be observable at feasible optical intensities.

The paper is organized as follow. In Section II we discuss epsilon-near-zero nonlinear metamaterials, we introduce the basic theory for investigating
bright solitons and we derive their existence conditions even for values of the nonlinear parameters not available in nature but available through
the proposed nonlinear metamaterial. In Section III and Section IV we consider two-peaked and flat-top solitons, respectively, and we focus on their
own underlying self-trapping supporting mechanisms by discussing their mutual differences and the role played by the unconventional values of the
nonlinear parameters. In Section V we propose a feasible and realistic composite medium (nonlinear metamaterial) suitable for observing two-peaked
and flat-top solitons. Finally, in Sec. VI, we conclude with a discussion and summary of our results.

\section{Bright Solitons in nonlinear epsilon-near-zero metamaterials}

Consider monochromatic transverse magnetic (TM) fields $\textrm{E} = \left[{\bf E} \exp {(-i \omega t)} \right]$, $\textrm{B}=Re \left[{\bf B}
\exp{(-i \omega t)} \right]$ (of frequency $\omega$), where
\begin{eqnarray} \label{TM-field}
{\bf E} &=& E_x(x,z) \hat{\bf e}_x+ E_z(x,z) \hat{\bf e}_z, \nonumber \\
{\bf H} &=&  H_y(x,z) \hat{\bf e}_y,
\end{eqnarray}
propagating through a nonlinear metamaterial whose effective medium response is described by the constitutive relations
\begin{eqnarray} \label{H-D}
&& {\bf D}= \epsilon_0 \epsilon {\bf E}+\epsilon_0 \chi \left[ |{\bf E}|^2 {\bf E} +\gamma ({\bf E} \cdot {\bf E}) {\bf E}^* \right], \nonumber \\
&& {\bf B}=\mu_0 \mu {\bf H},
\end{eqnarray}
where $\epsilon$ and $\mu$ are the linear dielectric permittivity and magnetic permeability, respectively, whereas $\chi$ and $\gamma$ are parameters
characterizing the overall effective nonlinear Kerr response. As opposed to conventional media where $\epsilon \geq 1$, $\mu \simeq 1$ and $\gamma$
assumes three values depending on the physical mechanism supporting the Kerr nonlinear response \cite{Boyd1} (i.e., $\gamma = 0$ for
electrostriction, $\gamma = 0.5$ for nonresonant electronic response and $\gamma = 3$ for molecular orientation), we hereafter focus on suitable
metamaterials designed to exhibit a very small dielectric permittivity $|\epsilon| \ll 1$ \cite{Engheta}, an arbitrary $\mu$ and with $\gamma$
attaining, in principle, any real value. The possibility of tailoring both linear and nonlinear medium responses has been investigated in the
pioneering paper by J. E. Sipe et al. of Ref.\cite{Sipe}  where the authors consider a composite structure comprised of spherical inclusion particles
embedded in a host material. Remarkably, if the inclusions respond linearly and the host material responds nonlinearly, the composite is
characterized by a nonlinear effective Kerr response whose parameter $\gamma$ can span the range $0<\gamma<3$ \cite{Sipe}. Recently, A. Ciattoni et
al. \cite{Ciatt1} have improved these results by showing that TM optical waves travelling through a nonlinear layered medium, in the long wavelength
limit, experience a nonlinear optical response described by Eqs.(\ref{H-D}) whose electromagnetic parameters can be independently tailored. In
particular, using positive and negative (linear and nonlinear) dielectric layers, they have shown that a suitable composite tailoring allows $\gamma$
to achieve any prescribed value (even encompassing negative values) and $\epsilon$, at the same time, to assume a value very close to zero. The
evident consequence of operating with a very small dielectric permittivity is that the extreme nonlinear regime can be observed where the nonlinear
polarization can be comparable with the linear part \cite{Ciatt1}, i.e. the condition
\begin{equation} \label{extreme}
\chi \left[ |{\bf E}|^2 {\bf E} +\gamma ({\bf E} \cdot {\bf E}) {\bf E}^* \right] \approx  \epsilon {\bf E}
\end{equation}
can be satisfied at small and feasible optical intensities. Note that in standard media where $\epsilon \approx 1$, condition of Eq.(\ref{extreme})
is strictly meaningless since the Kerr nonlinearity is always a perturbation to the linear polarization and, for very high intensities, it always
displays a form of saturation thus departing from the cubic behavior and never reaching values of the order of $\epsilon {\bf E}$. Therefore, even
though Eqs.(\ref{H-D}) formally coincide with the standard nonlinear Kerr response, the considered media host a marked and hitherto unexplored
nonlinear behavior (extreme nonlinear regime) accessible at feasible optical intensities. It is worth stressing that, since the very small value of
the permittivity is here achieved by averaging negative and positive dielectric permittivities, it is evident that suitable gain media have to be
inserted among the underlying medium constituents in order to compensate losses due to the presence of negative dielectrics (see Sec. V).

We focus on non-diffracting solitary waves propagating along the $z$-axis of the kind
\begin{eqnarray} \label{waveguide}
&& E_x = \sqrt{|\epsilon/\chi|} u_x(\xi)\exp{(i \beta \zeta)}, \nonumber \\
&& E_z = i  \sqrt{|\epsilon/\chi|} u_z(\xi)\exp{(i \beta \zeta)},
\end{eqnarray}
where $\xi = \sqrt{|\mu \epsilon|} k_0 x$, $\zeta = \sqrt{|\mu \epsilon|} k_0 z$ (dimensionless spatial coordinates), $k_0=\omega/c$ and we require
both $\beta$ (dimensionless propagation constant) and $u_x$ and $u_z$ (dimensionless field amplitudes) to be real quantities. Note that, using
dimensionless field amplitudes introduced through Eqs.(\ref{waveguide}), the extreme nonlinear regime condition of Eq.(\ref{extreme}) can be stated
as
\begin{eqnarray}
u_x^2 & \approx & 1, \nonumber \\
u_z^2 & \approx & 1.
\end{eqnarray}
Maxwell equations $\nabla \times {\bf E}=i \omega {\bf B}$, $\nabla \times {\bf H}=-i \omega {\bf D}$ for the fields of Eqs.(\ref{TM-field}) and
Eqs.(\ref{waveguide}) and with the constitutive relations of Eqs.(\ref{H-D}) yield, after some algebra, the system of first order differential
equations
\begin{eqnarray} \label{syst2}
&& \beta \frac{d u_z}{d \xi} =\left[\beta^2- \sigma_\mu \epsilon_x^{(NL)}\right] u_x, \nonumber\\
&& \beta \frac{d u_x}{d \xi} = \frac{\beta^2 \epsilon_z^{(NL)} u_z -\left( \beta^2- \sigma_\mu \epsilon_x^{(NL)} \right)  \displaystyle
\frac{\partial \epsilon_x^{(NL)} }{\partial u_z} u_x^2 }{ \epsilon_x^{(NL)}+ \displaystyle \frac{\partial \epsilon_x^{(NL)}}{\partial u_x} u_x },
\nonumber \\
\end{eqnarray}
where $\sigma_\epsilon=\epsilon \chi/|\epsilon \chi|$, $\sigma_\mu=\mu \chi/|\mu \chi|$ and we have defined
\begin{eqnarray} \label{epNL}
\epsilon_x^{(NL)} &=&  \sigma_{\epsilon} + (1+\gamma) u_x^2+(1-\gamma)u_z^2,\nonumber\\
\epsilon_z^{(NL)} &=& \sigma_{\epsilon}+(1-\gamma) u_x^2+(1+\gamma)u_z^2,
\end{eqnarray}
as effective normalized nonlinear dielectric permittivities. We note that the system of Eqs.(\ref{syst2}) is equivalent to Maxwell's equations if the
condition
\begin{equation}
\label{ineq}  \epsilon_x^{(NL)}+ \displaystyle \frac{\partial \epsilon_x^{(NL)}}{\partial u_x} u_x \neq0,
\end{equation}
holds along the whole field profile. It is worth stressing that the system of Eqs.(\ref{syst2}) is integrable since it admits the first integral
\begin{eqnarray} \label{int}
&& F(u_x,u_z)=[(\beta^2-\sigma_\epsilon \sigma_\mu)u_x^2-\sigma_\epsilon \sigma_\mu u_z^2]\nonumber \\
&& -\frac{1}{2}\sigma_\mu (1+\gamma) (u_x^4+u_z^4) -\sigma_\mu (1-\gamma)u_x^2u_z^2 \nonumber \\
&& -\frac{1}{\beta^2} \big \{ (\beta^2-\sigma_\epsilon \sigma_\mu)-\sigma_\mu [(1+\gamma)u_x^2+(1-\gamma)u_z^2]\big\}^2u_x^2, \nonumber \\
\end{eqnarray}
i.e. the relation $dF/d\xi=0$ is satisfied along the solutions $u_x(\xi)$, $u_z(\xi)$ of Eqs.(\ref{syst2}) (see Ref.\cite{Ciatt1} for details). We
here investigate bright solitons whose normalized electric field components $u_x$ and $u_z$ asymptotically vanish at infinity and are spatially even
($u_x(\xi)=u_x(-\xi)$) and odd ($u_z(\xi)=-u_z(-\xi)$), so that we consider the boundary conditions $u_{x}(0)=u_{x0}$, $u_z(0)=0$, $u_{x}(+\infty)=0$
and $u_{z}(+\infty)=0$. The vanishing of the field at infinity implies that the first integral $F$ has to vanish along the whole soliton profile,
i.e. $F(u_x(\xi),u_z(\xi))=0$, a relation which, evaluated at $\xi=0$, allows us to analytically evaluate the propagation constant
\begin{equation}
\label{beta-formula} \beta^2=2\sigma_\mu\frac{[\sigma_\epsilon+(1+\gamma)u_{x0}^2]^2}{2\sigma_\epsilon+3(1+\gamma)u_{x0}^2}.
\end{equation}
Remarkably, bright solitons are represented in the phase space $(u_x,u_z)$ by the special level curves $F(u_x,u_z)=0$ which are orbits of the
dynamical system of Eqs.(\ref{syst2}) (known as homoclinic orbits) joining the equilibrium point $(u_x,u_z)=(0,0)$  to itself. The main consequence
of this identification is that bright solitons asymptotically coincide for $\xi \rightarrow \pm \infty$ with the linear inhomogeneous transverse
plane waves $e^{\mp\sqrt{\beta^2-\sigma_\epsilon\sigma_\mu} \xi}$, respectively, so that the relation
\begin{equation}
\label{beta} \beta^2-\sigma_\epsilon\sigma_\mu>0
\end{equation}
is a necessary condition for soliton existence. The existence of at least one turning point (i.e. where $du_z/d\xi=0$ occurs) along the orbit
corresponding to a bright soliton is an additional condition to be met and it is fulfilled by requiring the curves $F(u_x,u_z)=0$ and $[ \beta^2 -
\sigma_\mu \epsilon_x^{(NL)} ] u_x=0$ (see the first of Eqs.(\ref{syst2})) to have an intersection. These necessary conditions for the existence of
bright solitons, together with Eq.(\ref{ineq}), allow us to obtain the results of Table I where we report, for each possible combinations of
$\sigma_{\epsilon}$ and $\sigma_\mu$, the range of $u_{x0}^2$ (depending on $\gamma$) where bright solitons can be excited.
\begin{figure}[htbp]
\begin{center}
\includegraphics[width=1\linewidth]{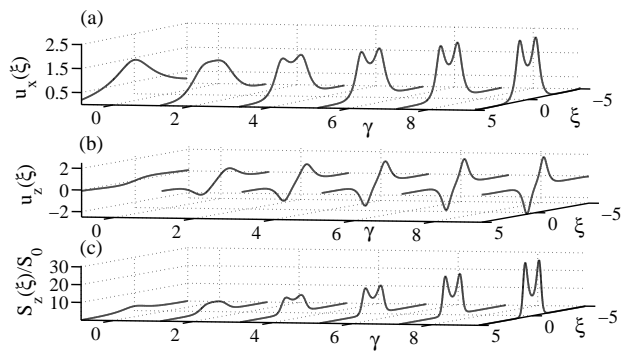}
\caption{Profiles of $u_x(\xi)$ (panel (a)), $u_z(\xi)$ (panel (b)) and normalized Poynting vector $S_z(\xi)/S_0$ (panel (c)) of various bright
solitons with $\sigma_\epsilon=1$, $\sigma_\mu=1$, $sign(\chi)=1$, $u_{x0}^2=2$ and $\gamma$ spanning the range $-0.8<\gamma<10$.}
\end{center}
\end{figure}
\begin{table}
  \begin{tabular}{ l  c  r  } \hline \hline
                & $\sigma_\epsilon=1, \sigma_\mu=1$  & $\sigma_\epsilon=-1, \sigma_\mu=-1$ \\ \hline
    $\gamma>-1$ &  $ \forall u_{x0}^2$                 &                                      \\ \hline
    $\gamma<-1$ &                                    & $ u_{x0}^2<\alpha(\gamma)$                  \\ \hline \hline
  \end{tabular}
\label{table1} \caption{Bright soliton existence ranges of $u_{x0}^2$ depending on $\gamma$, $\sigma_\epsilon$, $\sigma_\mu$. The parameter $\alpha$
is defined in Eq.(\ref{alfa}).}
\end{table}

\section{Two-peaked Solitons}

A first group of bright solitons (see Table I) exists for any values $u_{x0}^2$ in the range $\gamma>-1$ and for $\sigma_\epsilon=1, \sigma_\mu=1$
(i.e. for right-handed metamaterial with focusing nonlinearity, $\chi>0$, or for left-handed metamaterials with defocusing nonlinearity, $\chi<0$).
Note that this group of bright soliton encompasses the perfect optical solitons described in Ref.\cite{Ciatt4} where authors solely considered the
standard Kerr nonlinearity arising from the electronic nonresonant response ($\gamma=0.5$). In panels (a) and (b) of Fig.1 we report the bright
soliton profiles of $u_x$ and $u_z$, respectively, for $u_{x0}^2=2$, $-0.8<\gamma<10$ and $\sigma_\epsilon=1, \sigma_\mu=1$. For completeness, in
Fig.1(c) we report the profiles of the normalized Poynting vector $S_z(\xi)/S_0=sign(\mu) [(\beta u_x- du_z/d\xi)u_x]\hat {\bf e}_z $ (where
$S_0=\sqrt{(\epsilon_0|\epsilon|^3)/(4 \mu_0 |\mu| \chi^2)}$) corresponding to the solitons reported in Fig.1(a) and Fig.1(b).
\begin{figure}[htbp]
\begin{center}
\includegraphics[width=1\linewidth]{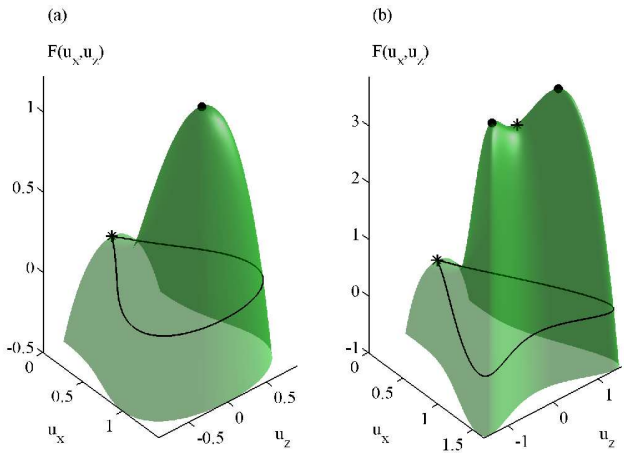}
\caption{(Color online) Plots of $F(u_x,u_z)$ evaluated for $\sigma_\epsilon=1$, $\sigma_\mu=1$, $u_{x0}^2=2$, $\gamma=0.5$ (panel (a)), $\gamma=3$
(panel (b)). Black lines correspond to the homoclinic orbits whereas the stars and the circles are located at the saddle points and maxima of
$F(u_x,u_z)$, respectively.}
\end{center}
\end{figure}
From Fig.1, we note that two-peaked bright solitons can be excited, i.e. solitons whose electric field $x$-components and Poynting vectors show two
pronounced peaks along their transverse profiles, the higher the value of $\gamma$ the more pronounced the two peaks. In order to grasp the
underlying mechanism producing such an atypical (for the Kerr nonlinearity) two-peaked soliton profile, we consider the equilibrium points of the
system of Eqs.(\ref{syst2}) (i.e. $du_x/d\xi=0$ and $du_z/d\xi=0$ and hence necessarily coinciding with the critical points of the first integral
$F(u_x,u_z)$) which, in addition to the origin $(u_x,u_z)=(0,0)$, are given by the relations
\begin{eqnarray} \label{max_cond}
&&\beta^2- \sigma_\mu \epsilon_x^{(NL)}=0, \nonumber \\
&&\epsilon_z^{(NL)} u_z=0.
\end{eqnarray}
Note that the number of the critical points of $F$ depends, in addition to $u_{x0}^2$, on the value of the nonlinear parameter $\gamma$. In fact, for
$|\gamma|<1$ (see Fig.2(a)) $F$ has only a maximum at the point $(u_x,u_z)=(\sqrt{(\beta^2-1)/(1+\gamma)},0)$ (belonging to the axis $u_z=0$) and a
saddle point at the origin $(u_x,u_z)=(0,0)$; on the other hand, in the case $\gamma > 1$ and $u_{x0}^2 > \delta$ (see Fig.2(b)) where
\begin{equation}
\delta(\gamma) = \frac{(\gamma+2) + \sqrt{5\gamma^2+2 \gamma}}{2(\gamma^2-1)},
\end{equation}
$F$ has two saddle points (located at $(u_x,u_z) = (\sqrt{(\beta^2-1) /(1+\gamma)},0)$ and the origin $(u_x,u_z)=(0,0)$) and two maxima not belonging
to the axis $u_z=0$ (and hence characterized by the condition $\epsilon_z^{(NL)} = 0$, see Eqs.(\ref{max_cond})). From Fig.2 it is evident that, in
the considered situations, the homoclinic (solitonic) orbits make a turn around the maxima of the function $F(u_x,u_z)$ so that the shape of the
orbits (and consequently the soliton profiles) depends on the number of maxima enclosed by the homoclinic loop. Therefore, since there are one and
two maxima for $|\gamma| <1$ and $\gamma >1$ (for $u_{x0}^2 > \delta$), respectively, we conclude that homoclinic orbits undergo a qualitative
structural change at $\gamma=1$. Correspondingly, for $|\gamma| < 1$ homoclinic orbits are simple loops (see Fig.2(a)) and solitons are bell-shaped
whereas for $\gamma > 1$ (for $u_{x0}^2 > \delta$) homoclinic orbits are structured loops (see Fig.2(b)) and two-peaked solitons occur. In other
words (in a language more close to nonlinear optics) the discussed two kinds of bright solitons are literally supported by two different
self-trapping mechanisms, and the nonlinear metamaterials employed in this paper provide the additional parameter $\gamma$ to switch from one
mechanism to the other. It is worth noting that the first of Eqs.(\ref{max_cond}) globally holds for all the considered maxima whereas the second of
Eqs.(\ref{max_cond}) is differently fulfilled since, for $|\gamma|<1$ and $\gamma>1$ (for $u_{x0}^2 > \delta$), maxima occur for $u_z=0$ and
$\epsilon_z^{(NL)}=0$, respectively. Therefore, bell-shaped soliton are not constrained by additional specific requirements and they actually
coincide with perfect solitons of Ref.\cite{Ciatt4}. On the other hand, two-peaked solitons supported by the novel self-trapping mechanism are due
(for $u_{x0}^2 > \delta$) to the existence of off-axis maxima of F corresponding to the condition $\epsilon_z^{(NL)}=0$. However, the normalized
effective nonlinear permittivity $\epsilon_z^{(NL)}$ can vanish only if the nonlinear contribution to the polarization $(1-\gamma)
u_x^2+(1+\gamma)u_z^2$ exactly balances the linear part $\sigma_\epsilon=1$ (see the second of Eqs.(\ref{epNL})) and it is worth repeating that this
condition (equivalent to Eqs.(\ref{extreme})) is realistically achievable solely by means of the nonlinear metamaterials with dielectric permittivity
very close to zero we are considering in this paper.

\section{Flat-top solitons}

A second group of bright solitons (see Table I) exists for $u_{x0}^2<\alpha$ where
\begin{equation} \label{alfa}
\alpha(\gamma) = \frac{ \sqrt{|\gamma|}-2 + \sqrt{5|\gamma| -4\sqrt{|\gamma|}}}{2(|\gamma|-1)(\sqrt{|\gamma|}+1)},
\end{equation}
in the range $\gamma < -1$ and for $\sigma_{\epsilon}=-1$ and $\sigma_{\mu}=-1$ (i.e for right-handed metamaterials with defocusing nonlinearity,
$\chi<0$, or for left-handed metamaterial with focusing nonlinearity, $\chi>0$).
\begin{figure}[htbp]
\begin{center}
\includegraphics[width=1\linewidth]{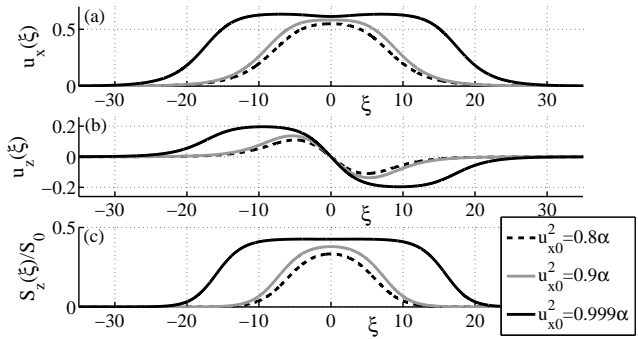}
\caption{Profiles of $u_x(\xi)$ (panel (a)), $u_z(\xi)$ (panel (b)) and normalized Poynting vector $S_z(\xi)/S_0$ (panel (c)) of various (with
different $u_{x0}^2$) bright solitons for $\sigma_\epsilon=-1$, $\sigma_\mu=-1$, $sign(\chi)=-1$ and $\gamma = -1.5$.}
\end{center}
\end{figure}
In Fig.3(a) and Fig.3(b) we report the soliton profiles (for various $u_{x0}^2$) of $u_x$ and $u_z$, respectively and in Fig.3(c) we plot the
corresponding normalized Poynting vector $S_z/S_0$, for $sign(\epsilon)=1$, $sign(\mu)=1$, $sign(\chi)=-1$, and $\gamma = -1.5$. Note that, for each
$\gamma < -1$, if $u_{x0}^2$ is very close to $\alpha$ (i.e. for $u_{x0}^2 > 0.99 \alpha$), flat-top bright solitons occur since their electric field
$x$-component and Poynting vector profiles are characterized by a flat and sharp core region (where the intensity is almost constant) which abruptly
stops being surrounded by lateral regions where the field approximately vanishes. On the other hand, for smaller intensities (i.e. for $u_{x0}^2 <
0.99 \alpha$) solitons are characterized by a bell-shaped profile.
\begin{figure}[htbp]
\begin{center}
\includegraphics[width=1\linewidth]{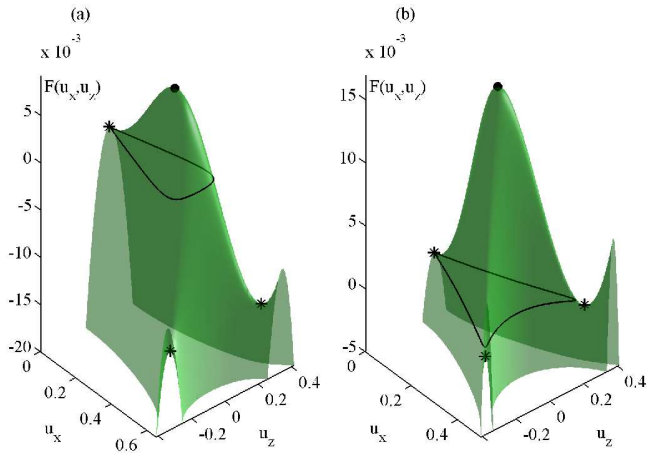}
\caption{(Color online) Plots of $F(u_x,u_z)$ evaluated for $\sigma_\epsilon=-1$, $\sigma_\mu=-1$, $\gamma=-7$, $u_{x0}^2=0.8 \alpha$ (panel (a)),
$u_{x0}^2=0.99 \alpha$ (panel (b)). Black lines correspond to the homoclinic orbits whereas the stars and the circles are located at the saddle
points and maxima of $F(u_x,u_z)$, respectively. }
\end{center}
\end{figure}

Even the mechanism leading to the formation of flat-top solitons can be understood by considering the critical point of the first integral $F$. For
$\gamma < -1$, it is simple to prove that the two off-axis critical points (i.e. with $u_z \neq 0$ and satisfying $\epsilon_z^{(NL)}=0$) obtained by
Eqs.(\ref{max_cond}) are saddle points of $F$ and this is particularly evident from Fig.4(a) and Fig.4(b) where we plot such saddle points (marked
with the stars) together with the first integral $F$ on the phase space $(u_x,u_z)$ for $u_{x0}^2 = 0.8 \alpha$ and $u_{x0}^2 = 0.99 \alpha$,
respectively (for the parameters $\sigma_\epsilon=-1$, $\sigma_\mu=-1$, $\gamma=-7$). Analogously to the case of the above discussed two-peaked
solitons, the off-axis critical points of $F$ have an incisive impact on the structure of the homoclinic orbits and, in the case $\gamma <-1$ we are
considering, the fact that they are saddle points implies that homoclinic orbit can not make a turn around them. Therefore, by varying $u_{x0}^2$
(for each $\gamma < -1$) the mutual position (and distance) between the homoclinic orbit and the saddle points can be changed (see Fig.4) to the
point that, for $u_{x0}^2$ very close to $\alpha$, the homoclinic orbit almost collides with the saddle points. When this happens, at the points
$(u_x,u_z)$ of the homoclinic orbit which are very close to the saddle points, the RHS of Eqs.(\ref{syst2}) are very small (since at the saddle they
exactly vanish) and therefore $d u_x /d\xi$ and $d u_z /d\xi$ are very small as well, thus explaining the obtained flat-top soliton behavior. In the
language of nonlinear optics, we can state that, for $\gamma <-1$, Kerr nonlinearity provides an additional self-trapping mechanism very different
from the above discussed two mechanisms supporting bell-shaped ($|\gamma|<1$) and two-peaked ($\gamma >1$) solitons. It is worth stressing again
that, since the two saddle points occur where the effective nonlinear permittivity vanishes ($\epsilon_z^{(NL)}=0$), the overall discussed flat-top
soliton phenomenology can be uniquely observed by means of the extreme nonlinear metamaterials we are considering in this paper.

\section{Nonlinear metamaterial design and soliton feasibility}
The possibility of observing both two-peaked and flat-top solitons is substantially based on the availability of nonlinear Kerr media characterized
by a very small dielectric permittivity and a parameter $\gamma$ in the ranges $\gamma < -1$ and $\gamma > 1$, respectively. In order to discuss the
feasibility of metamaterials exhibiting such unconventional properties, consider the composite structure reported in Fig.5 and consisting of
alternating, along the $y-$ axis, two kind of nonmagnetic layers of thicknesses $d_1$ and $d_2$, respectively. We here consider monochromatic
radiation whose free-space wavelength is $1550 \: nm$ and whose electromagnetic TM field geometry is also reported in Fig.5 from which it is evident
that such an electromagnetic field experiences an isotropic response. The layers of kind $1$ are filled with a suitable semiconductor which can be
doped to lower its plasma frequency to the point that, at the considered wavelength, its dielectric permittivity has a negative real part and a
considerably small imaginary part \cite{Gordon} (Zinc oxide is a good candidate that can be doped sufficiently high, for example with aluminium or
gallium, to satisfy these requirements \cite{Hirama}); besides, semiconductors are characterized by a standard nonresonant electronic Kerr response
(for which $\gamma = 0.5$). Therefore, for the layers of kind $1$, we can choose $\epsilon_1 = -8 + 0.1 i$, $\chi_1 = -5 \cdot 10^{-20} m^2/V^2$ and
$\gamma_1 = 0.5$ \cite{Sofian,Kulyk}. The layers of kind $2$ have to be filled with a nonlinear medium whose Kerr mechanism is different from the
nonresonant electronic one (otherwise the whole structure would exhibit $\gamma = 0.5$) and they have to host a gain mechanism for compensating
losses produced by the semiconductor (negative dielectric) layers of kind $1$.
\begin{figure}[htbp]
\begin{center}
\includegraphics[width=1\linewidth]{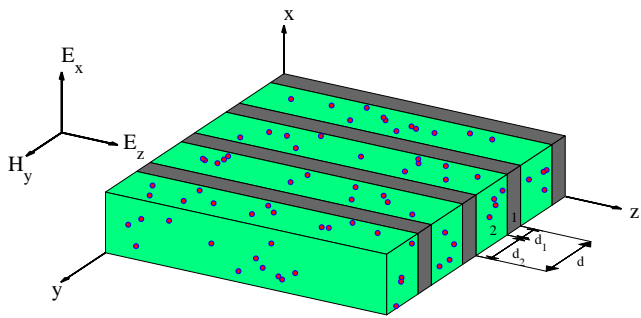}
\caption{(Color online) Composite nonlinear metamaterial for achieving $\epsilon \ll 1$ and $\gamma>1$ or $\gamma<-1$. The layers are staked along
the magnetic field direction of the impinging TM wave. Layers of kind $1$ are filled with a doped semiconductor (as ZnO:Al or ZnO:Ga) whose doping is
tuned to largely reduce the plasma frequency and hence to provide a negative dielectric with small absorption. Layers of kind $2$ are filled with a
nonlinear medium characterized by molecular orientation Kerr response and hosting dispersed semiconductor quantum dots.}
\end{center}
\end{figure}
Therefore, for the layers of kind $2$, we choose a background medium for which $\epsilon_b = 2.66$, $\chi_b =3 \cdot 10^{-20} m^2/V^2$ and $\gamma_b
= 3$ (the nonlinear parameters being typical of media characterized by molecular orientation Kerr response \cite{Boyd1}), hosting semiconductor
nanoparticles (optically pumped quantum dots) for which $\epsilon_{QD} = 11.8 -5.7 i$, $\chi_{QD} = 10^{-20} m^2/V^2$ and $\gamma_{QD} = 0.5$ (these
parameters characterizing suitable InAs/GaAs quantum dots as discussed in Ref.\cite{Bratk}). In order to obtain the effective electromagnetic linear
and nonlinear parameters of the whole structure of Fig.5, we exploit a two-step homogenization approach \cite{Cabuz} where, first, the effective
parameters of layers of kind $2$ are evaluated from the background medium and quantum dots properties and, second, the effective parameters of the
whole structure are obtained from those of the layers of kinds $1$ and $2$. Layers of kind 2 can be homogenized by exploiting standard Maxwell
Garnett approach (for evaluating the effective linear permittivity) and its nonlinear extension proposed by J.E. Sipe and R. Boyd in Ref.\cite{Sipe}
(for deducing the effective nonlinear parameters of the composite) so that, choosing the macroscopic volume filling fraction of the quantum dots
$f=0.03$, we obtain for the layers of kind $2$, $\epsilon_2 = 2.79 - 0.03 i$, $\chi_2 = \left( 3.61 \cdot 10^{-20} - 1.6 \cdot 10^{-21} i\right)
m^2/V^2$ and $\gamma_2 = 2.68 + 0.07 i$. If the period $d = d_1 + d_2$ (see Fig.5) is much smaller than the wavelength, effective linear
\cite{CaiSha} and nonlinear \cite{Boyd2} parameters can be introduced to describe the overall layered composite electromagnetic response so that,
choosing $d_1 = 52 \: nm$ and $d_2 = 148 \: nm$, we obtain $\epsilon = 1.5 \cdot 10^{-3} + 10^{-4} i$, $\chi = \left(1.38 \cdot 10^{-20} - 1.21 \cdot
10^{-21} i \right) m^2/V^2$ and $\gamma = 4.70 + 0.32 i$ (where the nonlinear parameters have been evaluated exploiting the approach of
Ref.\cite{Ciatt1}). Note that $Re(\epsilon) = 0.0015 \ll 1$, $Re(\gamma) = 4.70 > 3$ (different both from the standard values $0,0.5,3$ and those
achieved in Ref.\cite{Sipe}) and, remarkably, that all the imaginary parts can be neglected so that the conditions for observing the above discussed
two-peaked solitons are met. It is worth stressing that the fulfilling of all these conditions has been achieved by exploiting both the number of
available constituent materials and, most importantly, the freedom of tuning the parameters characterizing the composite (semiconductor doping,
nanoparticle size and volume filling fraction, layers' thicknesses, etc.). A two-peaked soliton, propagating through the above considered medium and
characterized by $u_{x0}^2=0.5$, requires the realistic and feasible peak intensity $775 \: MW/cm^2$ to be excited, value which should be compared
with the unphysical peak intensity $1.93 \cdot 10^7 \: MW/cm^2$ required for exciting a similar (i.e. with $u_{x0}^2=0.5$) two-peaked  soliton trough
carbon disulfide (for which $\epsilon_{CS_2} = 2.65$, $\chi_{CS_2} =3 \cdot 10^{-20} m^2/V^2$ and $\gamma _{CS_2} = 3$). Therefore, the extremely
high nonlinear behavior of the considered two-peaked soliton would require, in conventional media, optical intensities so high to prevent their
observation both because the nonlinear response would depart from the simple Kerr model and, most importantly, because the very high intensity would
literally damage the sample.

If in the above discussed composite structure the nonlinear susceptibility of layers of kind $1$ is changed into $\chi_1 = - 2 \cdot 10^{-19}
m^2/V^2$, by following the above described two-step homogenization approach, we obtain for the overall composite medium $\epsilon = 1.5 \cdot 10^{-3}
+ 1 \cdot 10^{-4} i$, $\chi = \left(-2.5 \cdot 10^{-20} - 1.21 \cdot 10^{-21} i \right) m^2/V^2$ and $\gamma = -1.83 + 0.14 i$. Therefore the
dielectric permittivity is very small, the overall nonlinear character is defocusing (since $Re(\chi) <0$) and, most importantly, $Re(\gamma) < -1$
so that all the conditions required for observing flat-top solitons are met. A flat-top soliton propagating through the considered medium and
characterized by $u_{x0}^2 = 0.99 \alpha(\gamma) = 0.32$ requires the feasible optical intensity $119 MW/cm^2$ to be excited.

\section{Conclusion}
In conclusion we have shown that a medium characterized by the constituent relations of Eqs.(\ref{H-D}), with $|\epsilon| \ll 1$ and $\gamma$
assuming values different from the three available in nature (i.e. $0$, $0.5$ and $3$), is able to support the propagation of two-peaked (for $\gamma
> 1$) and flat-top (for $\gamma < -1$) solitons. These two families of electromagnetic (perfect) solitons are supported by two distinct self-trapping
mechanisms whose main difference from the standard self-focusing process lies in the availability of critical points where the effective nonlinear
response vanishes (condition made possible by the very small value of the linear permittivity), points able to deeply affect the soliton shape. Since
the whole predicted soliton phenomenology is based on unconventional values of $\gamma$ we have proposed a realistic and feasible composite
metamaterial exhibiting such exotic nonlinear properties. It is worth stressing that, as a consequence of the small value of the linear dielectric
permittivity, the intensities required for exciting the considered solitons (characterized by highly nonlinear and highly nonparaxial behaviors) are
physically accessible. Therefore the soliton benchmark considered in the present paper allows us to argue that the two novel self-trapping mechanisms
can be efficiently exploited for fine-tuning the shape of an optical beam, shape control which can even become active by inserting electro-optic or
liquid-crystal based constituents within the nonlinear metamaterial.

\end{document}